\documentclass[preprint]{aastex}

\newcommand{\lya}{Ly$\alpha$}
\newcommand{\ha}{H$\alpha$}
\newcommand{\hb}{H$\beta$}
\newcommand{\hii}{H {\sc ii}}
\newcommand{\nii}{[N {\sc ii}]}
\newcommand{\oii}{[O {\sc ii}]}
\newcommand{\oiii}{[O {\sc iii}]}
\newcommand{\kp}{$K^{\prime}$}
\newcommand{\degree}{$^{\circ}$}
\newcommand{\hst}{$HST$}
\newcommand{\msol}{$M_{\odot}$}

\shorttitle{Nature of SMM J14011+0252}
\shortauthors{Motohara et al.}

\begin{document}

\title{Nature of a Strongly-Lensed Submillimeter Galaxy SMM J14011+0252$^{1}$}

\author{Kentaro Motohara\altaffilmark{2}, Tadafumi Takata\altaffilmark{3},
 Fumihide Iwamuro\altaffilmark{4}, Shigeru Eto\altaffilmark{4}, Takanori Shima\altaffilmark{4} ,
 Daisaku Mochida\altaffilmark{4}, Toshinori Maihara\altaffilmark{4},
 Kouichiro Nakanishi\altaffilmark{5}, 
 \& Nobunari Kashikawa\altaffilmark{6}}
\altaffiltext{1}{Based on data collected at Subaru Telescope, which is operated by the National Astronomical Observatory of Japan.}
\altaffiltext{2}{Institute of Astronomy, University of Tokyo, Mitaka, Tokyo 181-0015, Japan, kmotohara@ioa.s.u-tokyo.ac.jp}
\altaffiltext{3}{Subaru Telescope, National Astronomical Observatory of Japan, Hilo, HI 96720}
\altaffiltext{4}{Department of Astrophysics, Kyoto University, Kyoto 606-8502, Japan}
\altaffiltext{5}{Nobeyama Radio Observatory, Nobeyama, Minamimaki, Nagano 384-1305, Japan}
\altaffiltext{6}{National Astronomical Observatory of Japan, 1-21-1, Mitaka, Tokyo 181-8588, Japan}

\begin{abstract}
We have carried out near-infrared $JHK$ spectroscopy of
a gravitationally lensed submillimeter galaxy SMM J14011+0252 at $z=2.565$, 
using OHS and CISCO on the Subaru telescope.
This object consists of two optical components, J1 and J2, which are lensed by the cluster Abell 1835.
J1 suffers additional strong lensing by a foreground galaxy at $z=0.25$ in the cluster.
The rest-frame optical \ha, \hb, and  \oii$\lambda$3727 lines are detected in both J1 and J2, 
and \nii$\lambda\lambda6548,6583$\ lines are also detected in J1.
A diagnosis of emission-line ratios shows that the excitation source of J1 is stellar origin, 
consistent with previous X-ray observations.
The continua of J1 and J2 show breaks at rest-frame 4000\AA, indicating relatively young
age. Combined with optical photometry, we have carried out 
model spectrum fitting of J2 and find that it is a very young ($\sim$50 Myr) galaxy of 
rather small mass ($\sim 10^8 M_{\odot}$) which suffers some amount of dust extinction.
A new gravitational lensing model is constructed to assess both magnification factor and 
contamination from the lensing galaxy of the component J1, using 
\hst-$F702W$ image.
We have found that J1 suffers strong lensing with magnification of $\sim 30$, and
its stellar mass is estimated to be $\lesssim 10^{9}\, M_{\odot}$.
These results suggest that SMM J14011+0252 is a major merger system at high redshift
that undergoes intense star formation, but not a formation site of a giant elliptical.
Still having plenty of gas, it will transform most of
the gas into stars and will evolve into a galaxy of $\lesssim 10^{10}\,M_{\odot}$.
Therefore, this system is possibly an ancestor of
a less massive galaxy such as a mid-sized elliptical or a spiral at the present.
\end{abstract}


\section{Introduction}
Since the discovery of a large population of Lyman break galaxies \citep[hereafter LBGs;][]{steidel03},
a question whether we are seeing the majority of high-$z$ stellar population
or more stars are hidden in dusty star-forming galaxies has been raised.
It is expected that massive galaxy formation at high-$z$ may
cause rapid chemical evolution and dust enrichment, 
leading them to a dusty phase like ultra-luminous infrared galaxies (hereafter ULIRGs) 
emitting most of their energy in the far-IR.
Such objects are supposed to suffer from severe dust extinction, 
be unable to be picked up as LBGs defined by 
a spectral break in the rest-frame UV and only be detected in the far-infrared to submillimeter wavelength.

Deep submillimeter surveys by SCUBA on JCMT have revealed a 
large population of dusty galaxies at high-$z$, whose far-IR
luminosity exceeds $10^{12}\,L_{\odot}$ satisfying the criterion of ULIRGs
\citep[see ][and references therein]{blain02}.
Their optical to near-IR (NIR) color is often red, and some of them display no optical/NIR
counterpart even using 8-10m telescopes \citep{ivison02, fox02, frayer04}.
X-ray observations reveal that some of them contain type-2 AGNs, 
but contribution to the total luminosity is not clear \citep{alexander03}.
Submillimeter to radio flux ratios 
indicate that these objects lie at $z>1$ \citep{lilly99, gear00, hall01, smail02, ivison02, fox02},
and spectroscopic follow-ups of radio-selected sources reveal their 
median redshift to be around 2.4 \citep{chapman03a}.
These objects are therefore thought to be an important population to probe the star-formation 
history of the universe.
Due to their red color and faintness, 
their spectroscopic observations are time-consuming,
so there is only a small sample of spectroscopically 
observed objects in the NIR \citep{ivison00, frayer03, smail03a, smail03b, tecza04, simpson04}.
However, spectroscopy in the NIR is important because they 
cover the rest-frame optical wavelength, which provides us important 
information of their physical status, ionizing photons and stellar population.
We therefore carried out low-resolution NIR spectroscopic observations of SMM J14011+0252, one of the 
brightest submillimeter galaxies at high-$z$.

SMM J14011+0252 is discovered behind the cluster Abell 1835 ($z=0.25$) as a weakly lensed 
object \citep{smail98, barger99, frayer99}.
Intrinsic far-IR luminosity of $7\times10^{12} L_{\odot}$ indicates that this is an ULIRG, 
and inferred star formation rate (hereafter SFR) is few$\times 10^3\,M_{\odot}$ yr$^{-1}$  \citep[][hereafter I00]{ivison00}. 
It consists of two optical/NIR components; J1 and J2. They both lie at the same redshift of $z=2.56$, and 
the separation is 2\farcs1.
Both are red in the rest-frame optical ($R-K=3.8$ for J1 and 3.2 for J2) and
J1 is also red in the rest-frame UV ($U-R=2.1$) while J2 is bluer ($U-R=1.1$).
Both show only weak \lya\ emission (I00). No other emission line is detected in the rest-frame UV, but
strong \ha+\nii$\lambda\lambda6548, 6583$ lines exist in the NIR spectra \citep[I00; ][]{tecza04}.
\hst -$F702W$ image reveals a more complex structure of J1; it consists of several bright knots
surrounded by an extended diffuse component with an ERO component (J1n) elongated toward north \citep{ivison01}.
Integral field spectroscopy in the $JHK$-bands discovers an extended \ha~ emission cloud toward 
J1n \citep{tecza04}. A continuum break at rest-frame 4000\AA~ is found in J1, which indicates recent ($\sim 200$Myr) 
massive star-formation activities.

Radio observations detect CO($3-2$) \citep{frayer99, ivison01, downes03} and CO($7-6$) \citep{downes03} lines 
at $z=2.565$, 
which imply strong star-formation activity. 
The detection of CO($7-6$) indicates that temperature of the gas cloud is $30-40$K and its density 
is $n_{\rm H_2}\sim2000$ cm$^{-3}$ \citep{downes03}. The CO emission is
extended over 2\farcs2$\times$$<$0\farcs5 and centered at J1 \citep{downes03}.
A deep X-ray observation by Chandra shows no sign of X-ray emission \citep{fabian00}, 
suggesting that this object doesn't harbor a luminous AGN.

In addition to above, a detailed re-analysis of the optical spectrum reveals several
absorption features at $z=0.25$ including
Ca H+K  \citep[Smail et al, private communication (2003); ][]{downes03}. This suggests that
the main component of J1 is a foreground evolved galaxy, possibly an elliptical galaxy 
in Abell 1835, and that the other knots of J1 are strongly-lensed images of the submillimeter galaxy at $z=2.565$.
\citet{downes03} estimated the magnification factor to be $\sim25$ from their CO observation.

Here, we report results of 
low-resolution NIR $JHK$ spectroscopy 
of J1 and J2,
using OH-airglow suppressor \citep[OHS;][]{iwamuro01} and Cooled Infrared Camera and 
Spectrograph for OHS \citep[CISCO;][]{motohara02} installed at the 
8.2m Subaru telescope \citep{iye04}.
Thanks to the high sensitivity of OH-airglow suppression technique combined with 
the 8m aperture, high-S/N continuum spectra from rest-frame 3000 to 6500\AA~ with several strong emission-lines
are obtained.

We describe our observations and data reductions in section 2, display results 
in section 3, construct a new gravitational lensing model and discuss 
implication to the nature of this object in section 4, and 
summarize them in section 5.
$H_0=71 \rm~ km~s^{-1} Mpc^{-1}$, $\Omega_{\rm M}=0.27$, and $\Omega_{\Lambda}=0.73$
are assumed throughout this paper.

\section{Observations and Data Reductions}

\subsection{$K$-Band Imaging and Spectroscopy}

 \kp -band imaging and $K$-band spectroscopy were carried out on 2002 February 28, 
with CISCO mounted at the IR-Nasmyth focus of the Subaru telescope.
The optical-Nasmyth secondary mirror was used, and the pixel scale was 0\farcs111 pix$^{-1}$.

 \kp -band images were taken with single exposure time of 20s and the telescope was nodded 
after every 3 frames by 10\arcsec\ in a 4-point dither pattern. A total of 12 frames was acquired,
for a total exposure time of 240s. The seeing size was 0\farcs 45.
Before the \kp-band imaging, a faint standard star FS 27 \citep{hawarden01} was observed for
flux calibration.

The $K$-band spectra were acquired with 0\farcs65 slit, providing wavelength resolution of 400 at 2.15\micron . 
The position angle of the slit was set at $-69$\degree\ to introduce both J1 and J2 on the slit.
Single exposure time was 200s, and the telescope was nodded 4\arcsec$-$6\arcsec\ after every 3 frames
for sky subtraction. A total of 24 frames was taken for a total exposure time of 4800s.
Then, A0 star SAO 120250 was observed for correction of the 
atmospheric and instrumental transmission curve.

\subsection{$H$-Band Imaging and $JH$-Band Spectroscopy}

 $H$-band imaging and $JH$-band spectroscopy were carried out on 2002 March 1, 2, and May 18,
with OHS and CISCO mounted at the IR-Nasmyth focus of the Subaru telescope.
The opt-Nasmyth secondary mirror was used, and the pixel scale was 0\farcs111 pix$^{-1}$.

 $H$-band images were taken with OHS in the imaging mode \citep[see ][]{iwamuro01}
before the spectroscopic observations.
Single exposure time was 50s and the telescope was nodded after every 3 frames by 
10\arcsec\ in the direction of the dark lane of OHS, which is the
region where the OH airglow is suppressed.
A total of 6 frames was acquired each day.
FS 135 was observed after the observations on March 1 and 2, and FS 141 on May 18 
for flux calibration.
The seeing size was 0\farcs7$-$0\farcs8.

 $JH$-band spectra were acquired using a 0\farcs95 slit, providing resolution of 
210 at 1.65\micron. 
Single exposure time was 1000s and the telescope was nodded after each exposure by 
10\arcsec\ along the slit.
The position angle of the slit was again set at $-69$\degree\ on March 1 and 2 to 
observe J1 and J2 simultaneously. On May 18, the position angle was set at 
24\degree\ to observe J1 and J1n.
Total exposure time was 6000s on March 1, 4000s on March 2, and 8000s on May 18.
F0 star SAO 120154, A0 star SAO 120250, and F5 star SAO 120186 were observed after
the target  on March 1, 2, and May 18 respectively, for correcting atmospheric and instrumental 
transmission curve.

The log of the spectroscopic observations are summarized in Table \ref{tab00}.

\subsection{Data Reductions}
The imaging data are reduced in a standard procedure of 
flat-fielding, sky subtraction, bad-pixel correction, residual sky subtraction, 
and shift-and-add. Photometry is performed in an aperture of same size as
the slit width and length used in the $JH$-band spectroscopy. 
The position angle of the aperture is set at $-69$\degree.
Measured flux is summarized in Table \ref{tab0}.

The spectroscopic data are
reduced in a standard procedure of flat-fielding, sky subtraction, bad-pixel correction, and
residual sky subtraction.
They are then corrected for the atmospheric and instrumental transmission curve, 
and are coadded to create final frames.
Spatial widths of the aperture to extract 1-D spectra are 2\farcs2 for J1 and 1\farcs7 for J2.
In the $K$-band, wavelength calibration is carried out using the airglow lines 
in the raw frames. In the $JH$-band, because almost all the airglow lines 
are eliminated significantly by OHS, wavelength calibration is done using a pixel-wavelength relation 
obtained by previous observations of CISCO without OHS where the airglow lines are available.
Systematic error of this pixel-wavelength relation is estimated to be less than
0.5 pixel ($<3$\AA). 
Flux of the spectra is calibrated by the photometry of the $H$- and \kp-band imaging in Table \ref{tab0}.

\subsection{\hst\ WFPC2 Imaging}
We also obtain archived \hst\ WFPC2 $F702W$ data of Abell 1835 field (PI: Kneib, J.-P; PID: 8249), 
to investigate morphology and optical magnitude.
All the frames with 2500s exposure each are combined into a single frame after cosmic-ray rejection.
Total exposure time is 7500s.

\section{Results}

\begin{figure}
\figurenum{1}
\epsscale{1}
\plotone{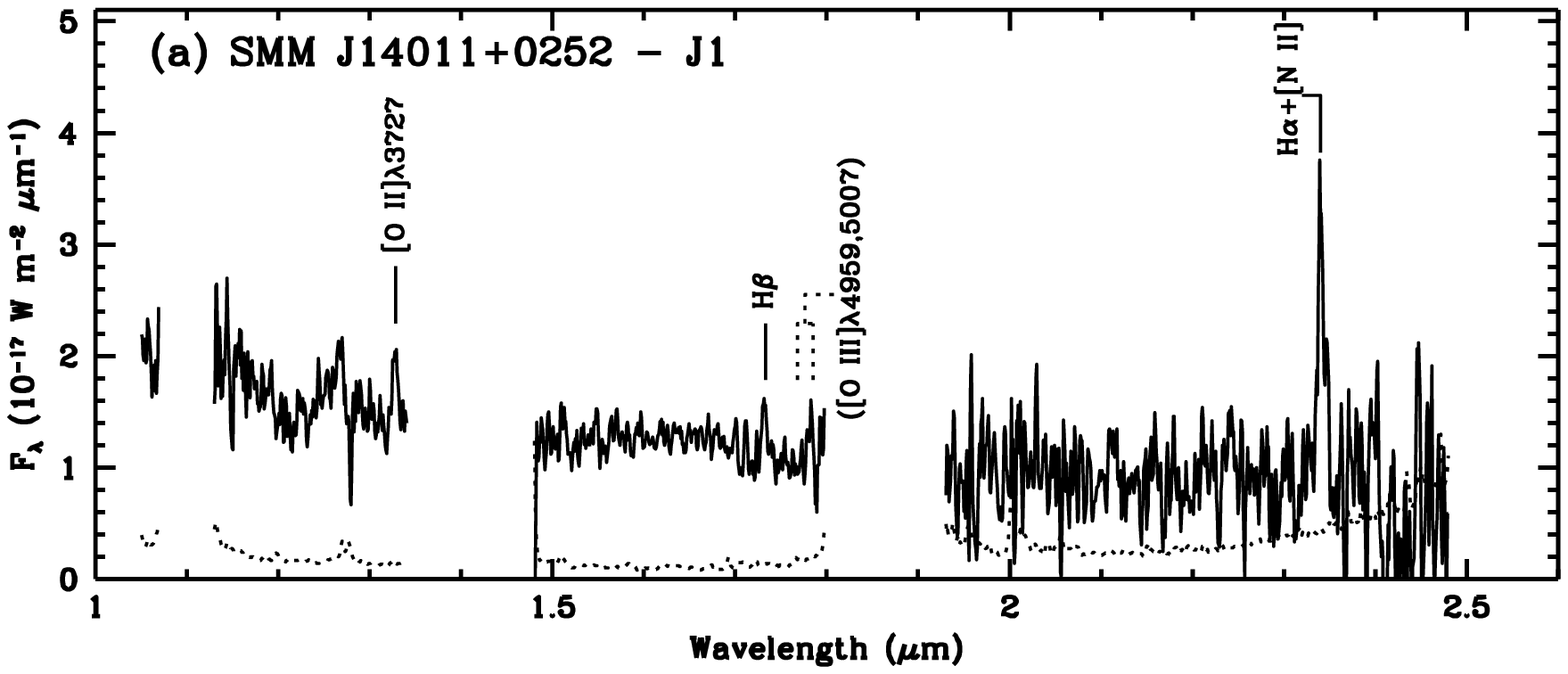}
\plotone{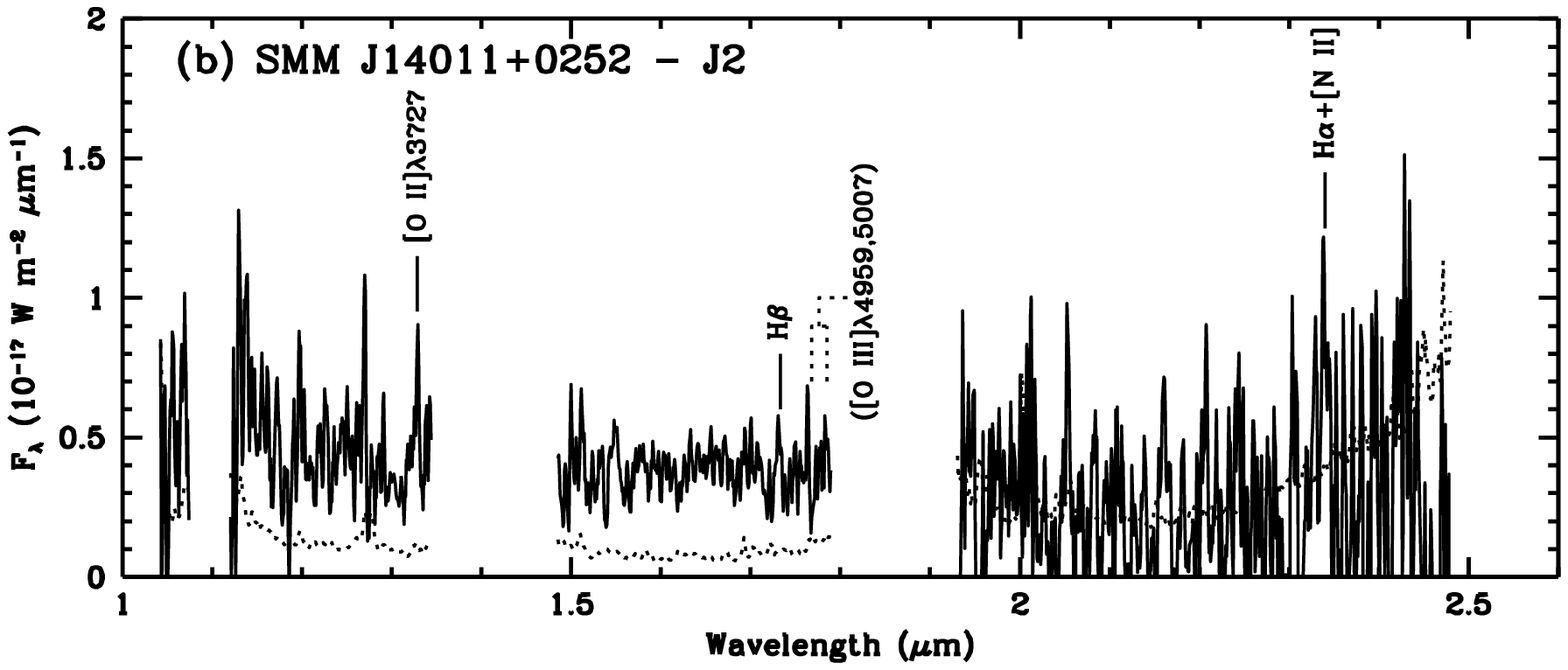}
\caption{ $JHK$ spectra of (a) SMM J14011+0252 J1 and (b) J2. Both spectra are 
smoothed by 3 pixels bin. Dotted lines indicate 1 $\sigma$ error level.}
\label{fig1}
\end{figure}

The final $JHK$ spectra of J1 and J2 extracted from the February and
March data are shown in Figure \ref{fig1}.
Although the wavelength resolution is lower ($\lambda/\Delta\lambda=200-400$) than that of
\citet{tecza04} ($\lambda/\Delta\lambda\sim 1500$), the quality of the continuum spectra is greater in the
$JH$-band, owe to the high sensitivity of OHS.
In J1, we detect strong \ha+\nii$\lambda\lambda 6548, 6583$ and relatively faint \oii$\lambda3727$ and \hb.
No \oiii$\lambda\lambda4959,5007$ is found. 
In J2, \ha, \hb, and \oii\ are detected, but all the lines are weaker than those in J1.
No line is resolved with the low resolution of 
OHS/CISCO, which is 750 km s$^{-1}$ in the $K$-band and 1400 km s$^{-1}$ in the $H$-band.
The $JH$-band spectrum of J1 taken in May is identical to that taken in March,
 with weak \hb\ and \oii\ without \oiii\ lines.
J1n is not detected.

\begin{figure}
\figurenum{2}
\epsscale{0.5}
\plotone{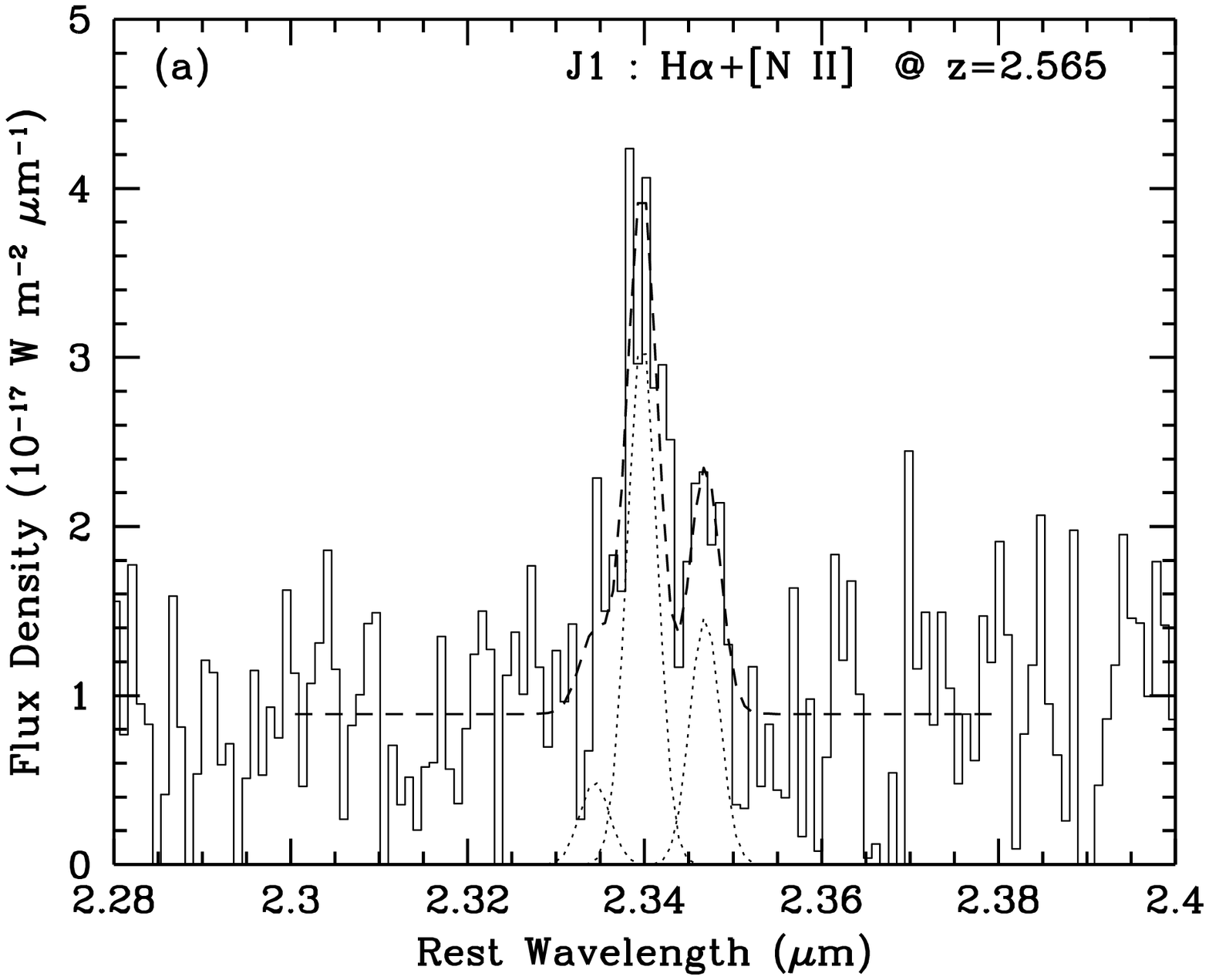}
\plotone{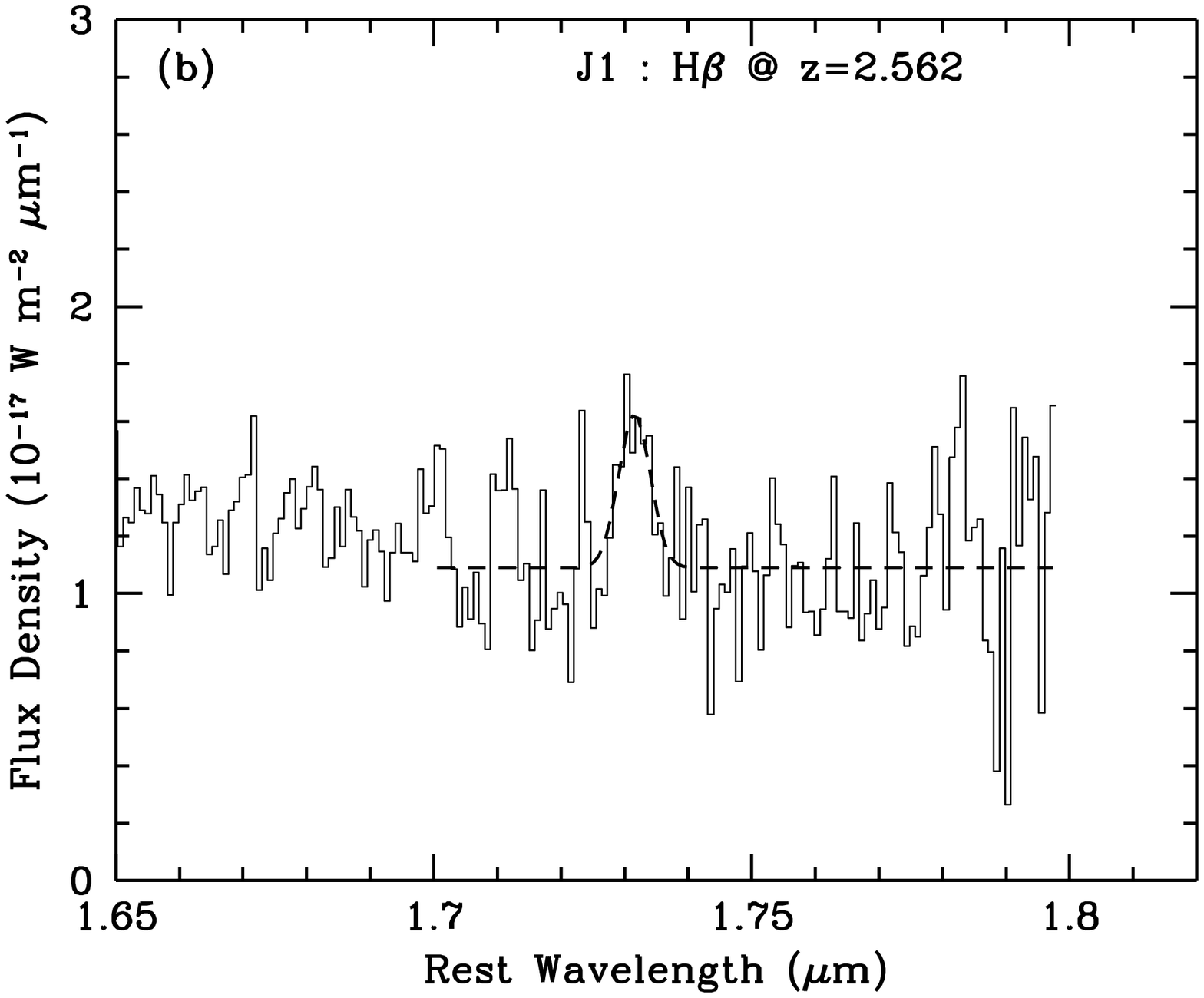}
\plotone{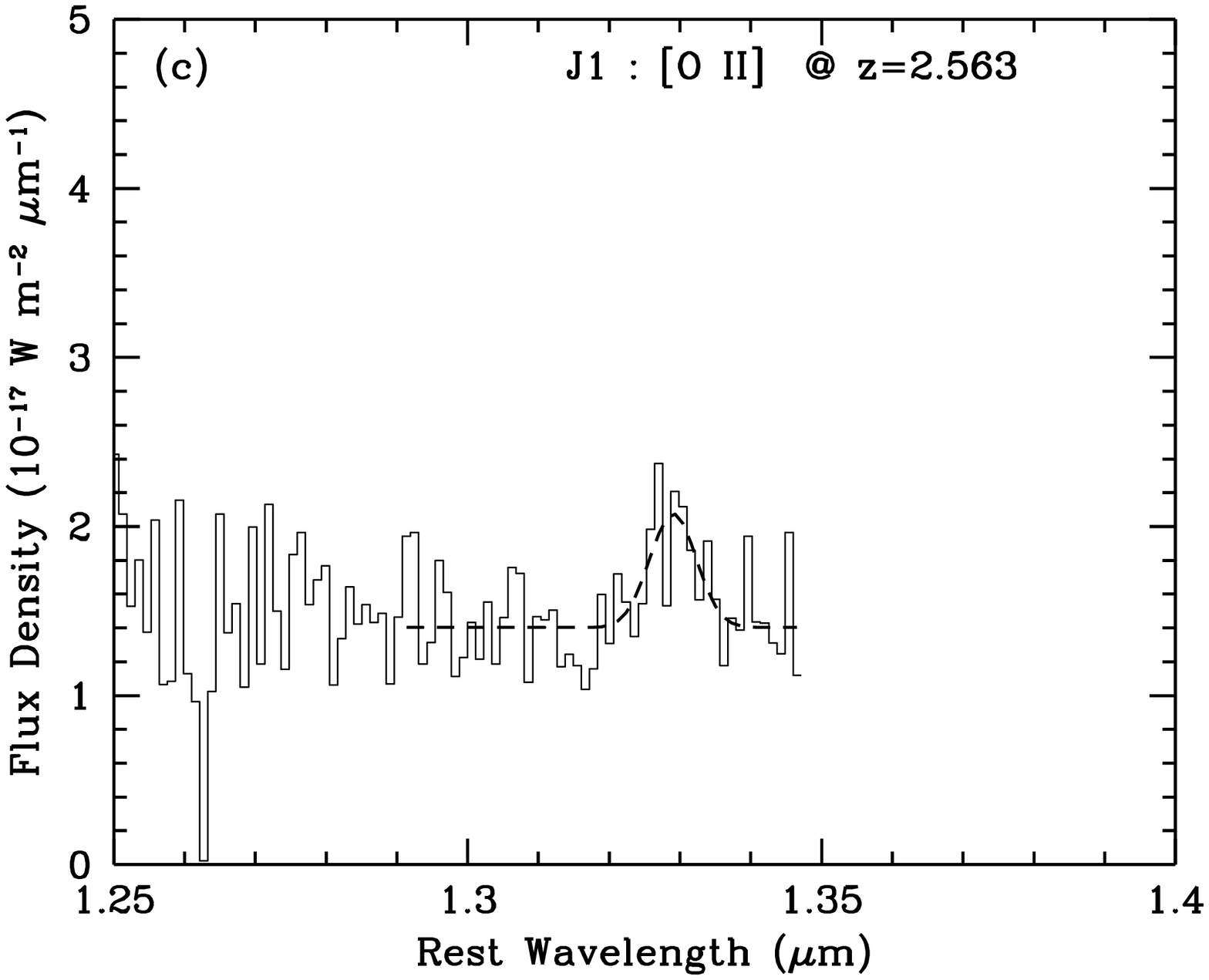}
\caption{(a) \ha+\nii, (b) \hb, and (c) \oii\ lines of J1, fitted with Gaussian profiles (dashed lines).}
\label{fig2}
\end{figure}

We next carry out line fitting with Gaussian profiles to estimate their
redshift, flux and equivalent width.
We also deblend \ha\ and \nii, which are merged due to the low resolution, assuming
\nii$\lambda$6583/\nii$\lambda$6548=3.0.
Because the S/N of the spectra is low, line widths are fixed to the resolving power 
determined by the slit width.
Results are given in Table \ref{tab1}, and 
fitted profiles of J1 are shown in Figure \ref{fig2}.
We also estimate an 1$\sigma$ upper limit of the line flux of \oiii$\lambda5007$ in J1 
to be $2.0\times10^{-20}$ W m$^{-2}$.
The lines of J2 are weaker, and the S/N is quite low. However, because the redshifts of all the lines match
within 1$\sigma$ errors, we conclude these detection to be real.

\section{Discussion}

\subsection{Emission Line Cloud of J1}
\begin{figure}
\figurenum{3}
\epsscale{1}
\plotone{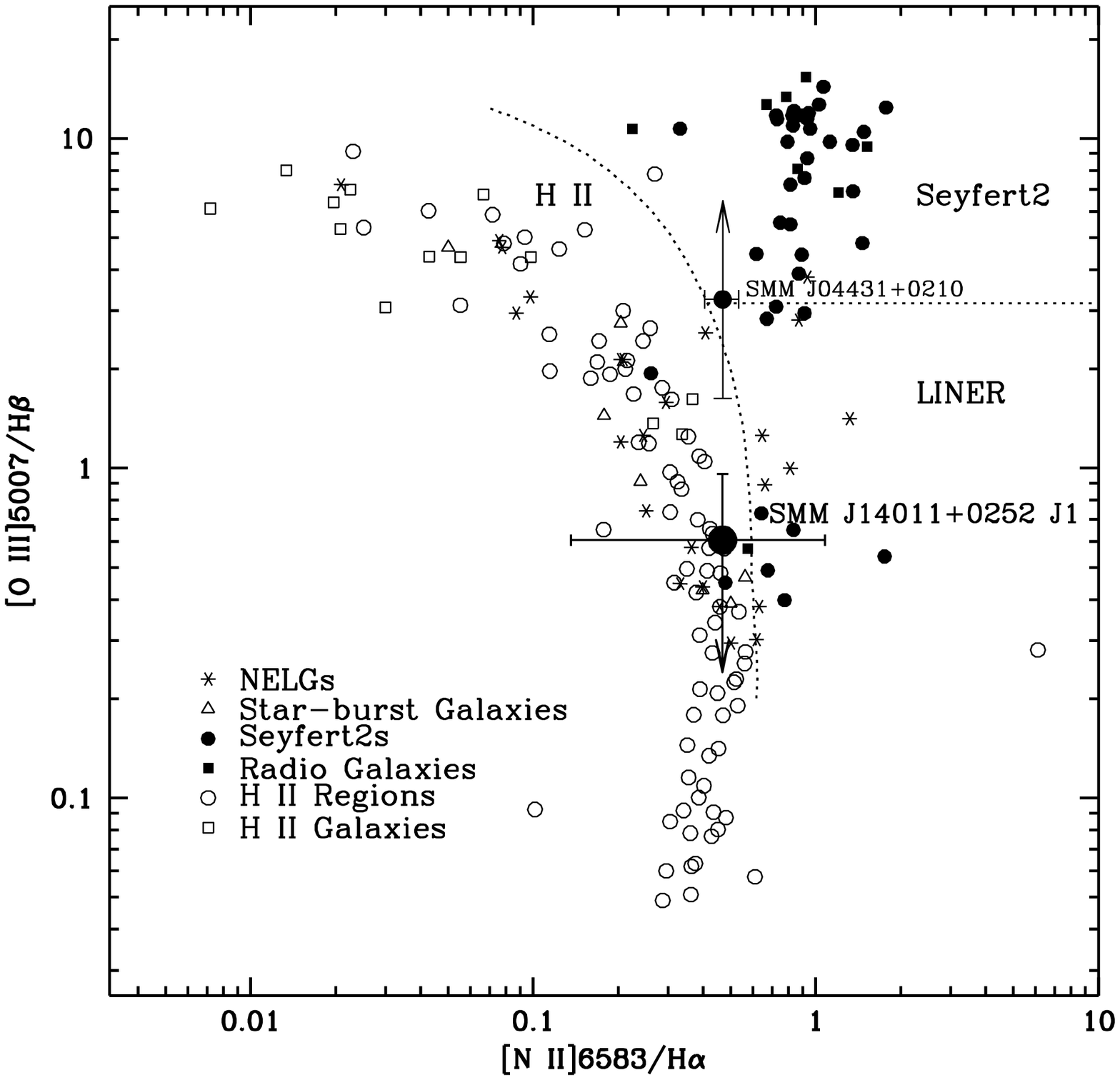}
\caption{Emission-line diagnostic diagram of \ha/\nii$\lambda6583$ vs. \oiii$\lambda5007$/\hb.
The large open circle with error bars is the line ratio of J1, while the large filled circle 
indicates that of SMM J04431+0210 taken from \citet{frayer03}.
The dotted lines show the separation between \hii\ regions, LINERs, and Type 2 AGNs taken from 
\citet{veilleux87} and \citet{filippenko92}.
Also the line ratios of Seyfert 2s(filled circles), radio galaxies(filled squares), narrow emission-line galaxies
(asterisks), \hii\ regions(open circles), 
\hii\ galaxies(open circles), and starburst galaxies(open triangles) taken from literatures
\citep{veilleux87, mccall85, french80, koski78, costero77} are plotted with smaller marks.}
\label{fig3}
\end{figure}

A diagram of \ha/\nii$\lambda6583$ and \oiii$\lambda 5007$/\hb\ is shown in Figure \ref{fig3}.
The line ratios 
indicate that the emission line cloud of J1 is ionized/excited by 
massive stars, not by a strong AGN, which is consistent with the 
X-ray observations 
showing no sign of a luminous AGN unless it is Compton thick \citep[][]{fabian00}.

Balmer decrement is measured to be \ha/\hb=4.0$\pm$2.2. 
If we assume the intrinsic 
line ratio of a star-forming region to be 2.9 \citep[Case B, $T=10^4$K;][]{osterbrock89}
and adopt the SMC extinction curve \citep{pei92},
dust extinction is estimated to be 
$E(B-V)=0.36^{+0.43}_{-0.36}$.
Due to the large errors in the line fluxes, the value is quite insecure.
Nevertheless, this indicates that the \hii\ region suffers extinction
of $A_V\sim1$.

Lower limit of the metallicity can be assessed by $O3N2$ or $R_{23}$  estimator.
The metallicity derived from $O3N2$ is $12+\log({\rm O/H})>8.7$ \citep[][]{pettini04}, while that from
$R_{23}$ is $12+\log({\rm O/H})>8.9$ \citep[][]{kobulnicky99}. 
These results are consistent with that of \citet{tecza04}, suggesting 
that this system is well polluted by metals supplied from the starburst activities.

\subsection{Gravitational Lensing Model}
\begin{figure}
\figurenum{4}
\epsscale{0.7}
\plotone{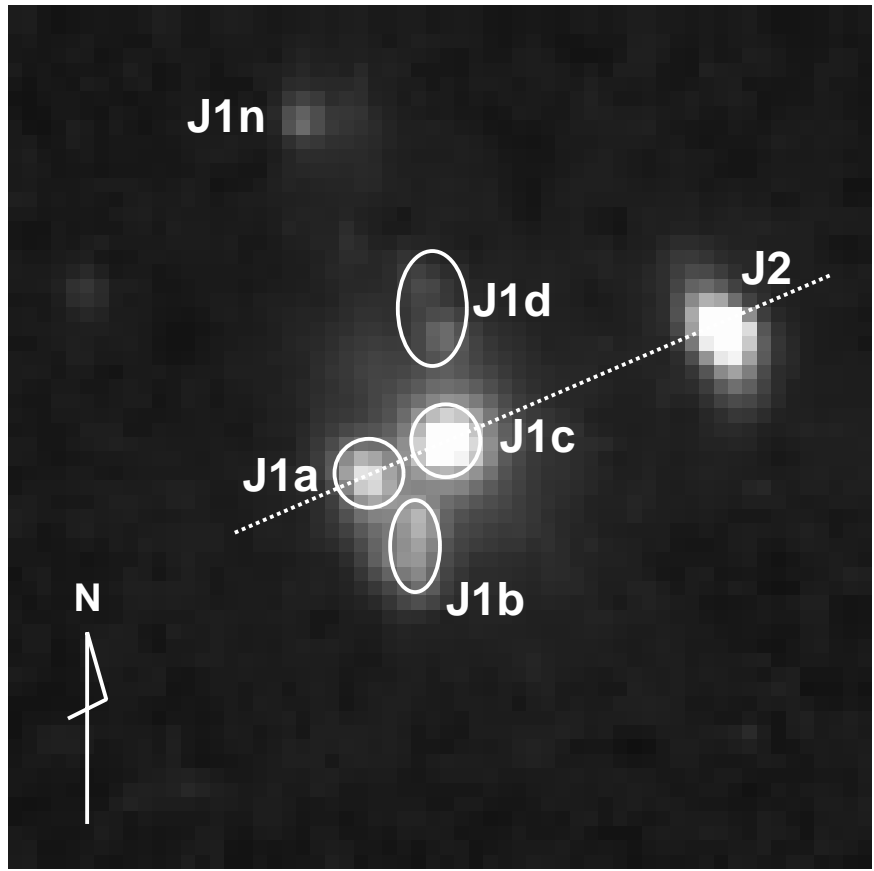}
\caption{$HST$-$F702W$ image of SMM J14011+0252. The dotted line 
indicates the slit position of the OHS/CISCO spectroscopy.
}
\label{fig4}
\end{figure}

The \hst-$F702W$ image is shown in Figure \ref{fig4}. It can be seen that J1 consists
of many sub-components. We name them to be J1a, J1b, J1c, and J1d as labeled 
in the figure.

Recently, Smail et al. (2003; private communication) found Ca H+K and other absorption lines 
at $z=0.25$ in the optical spectrum of J1, which suggests J1 to be contaminated and strongly lensed 
by a foreground galaxy. 
Also, as seen in the \ha~image (Figure 1 of \citet{tecza04}), the line brightness
becomes weak at the position of J1c and  a ``hole'' is seen there.
These results indicate that J1c is a member of Abell 1835 at $z=0.25$ and 
causing a strong lensing effect on J1 at $z=2.565$.

\citet{downes03} claimed that estimated brightness temperature of CO(3$-$2) 
is much smaller than that expected from the CO line ratio, and argued that this is 
due to the gravitational lensing effect.
They roughly estimate the magnification factor to be $\sim$25.
They also constructed a simple gravitational lensing model using the \hst-$F702W$ image,
assuming that J1a, J1b, J1n and J2 are all the lensed image of a single object.
However, this model failed to reproduce the position of the observed components.
In addition, it is known that the redshifts of J1 and J2 differ slightly \citep{ivison00, tecza04}
and they are thought to be different objects.
We therefore propose a new gravitational lensing model assuming that  
J1a, J1b, and J1d are lensed images of a single object by J1c and Abell 1835.
Hereafter, we refer to the lensing galaxy at $z=0.25$ as ``J1c'' 
and distinguish it from the lensed galaxy J1 at $z=2.565$.
 
\subsubsection{Lensing Model by GRAVLENS}
We use gravitational lensing software GRAVLENS \citep{keeton01}.
The model potential consists of two components J1c and Abell 1835, both lie at $z=0.25$.
The potential of J1c is assumed to be an isothermal ellipsoid with core radius $b=0$,
while that of Abell 1835 an NFW model with parameters taken from \citet{schmidt01}.
We then carry out fitting calculation of the positions and aperture
fluxes of J1a, J1b, and J1d, by varying velocity dispersion
$\sigma$, ellipticity $e$, and position angle of the ellipticity
$\theta_{e}$ of J1c.
All the parameters of the lensing model is shown in Table \ref{tab11}.
The uncertainty of each parameter is evaluated as follows:
We start at the best fit model, and vary a single parameter at once. 
For the varied model, $\chi^2$ fitting is carried out by varying the position of the source
to find the best-fit position.
The average displacement between the positions of the best-fit spots and the original 
spots in the image plane is then calculated,
and the error for the parameter is defined to be the value where the average displacement 
becmoes less than 0\farcs1.

Using this model, we also reconstruct the lensed image of J1n, J2, and the extended \ha~cloud 
found by \citet{tecza04}.
Positions and magnification factors of these components are listed in Table \ref{tab12}. 

\begin{figure}
\figurenum{5}
\epsscale{0.5}
\plotone{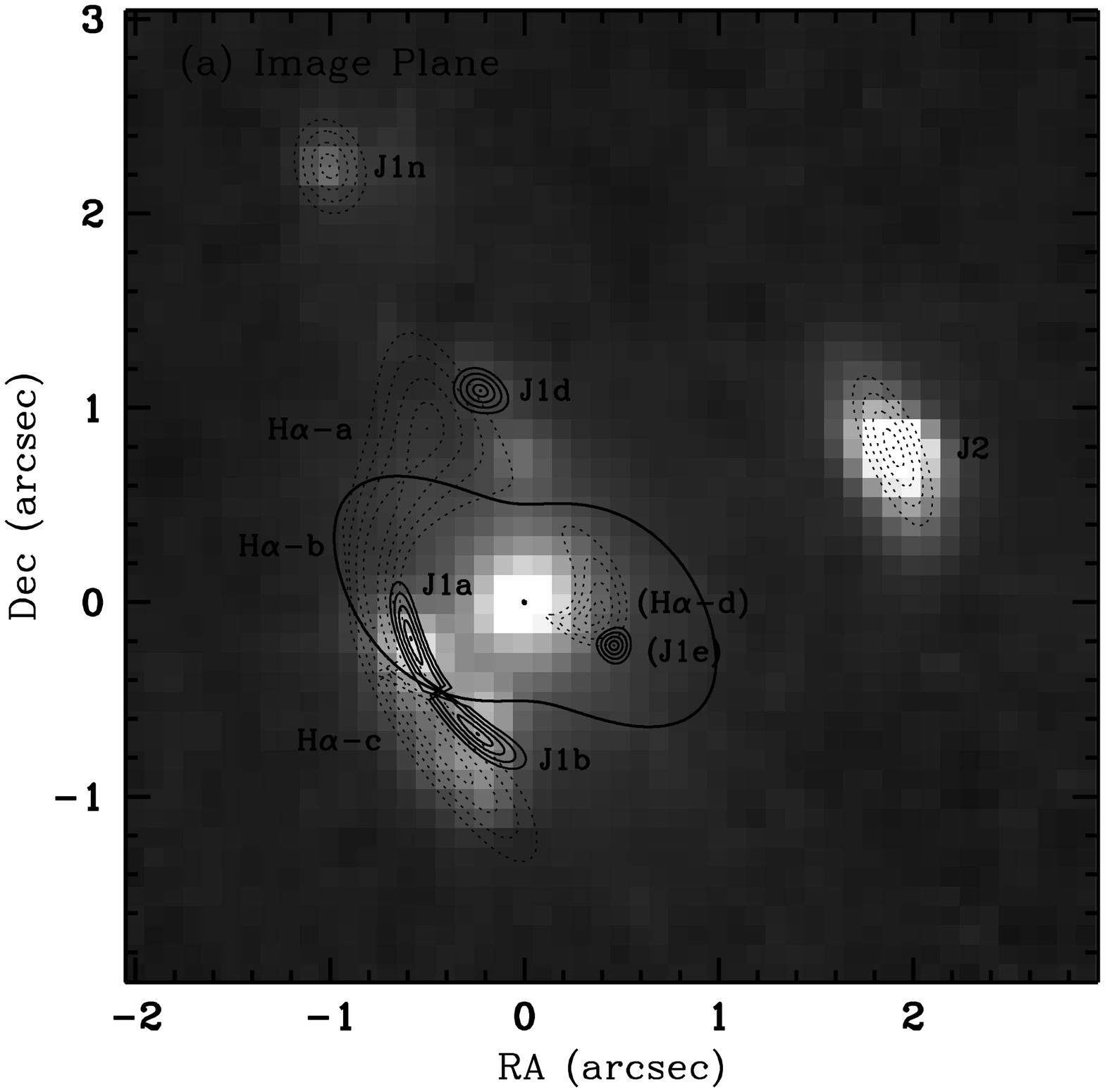}
\plotone{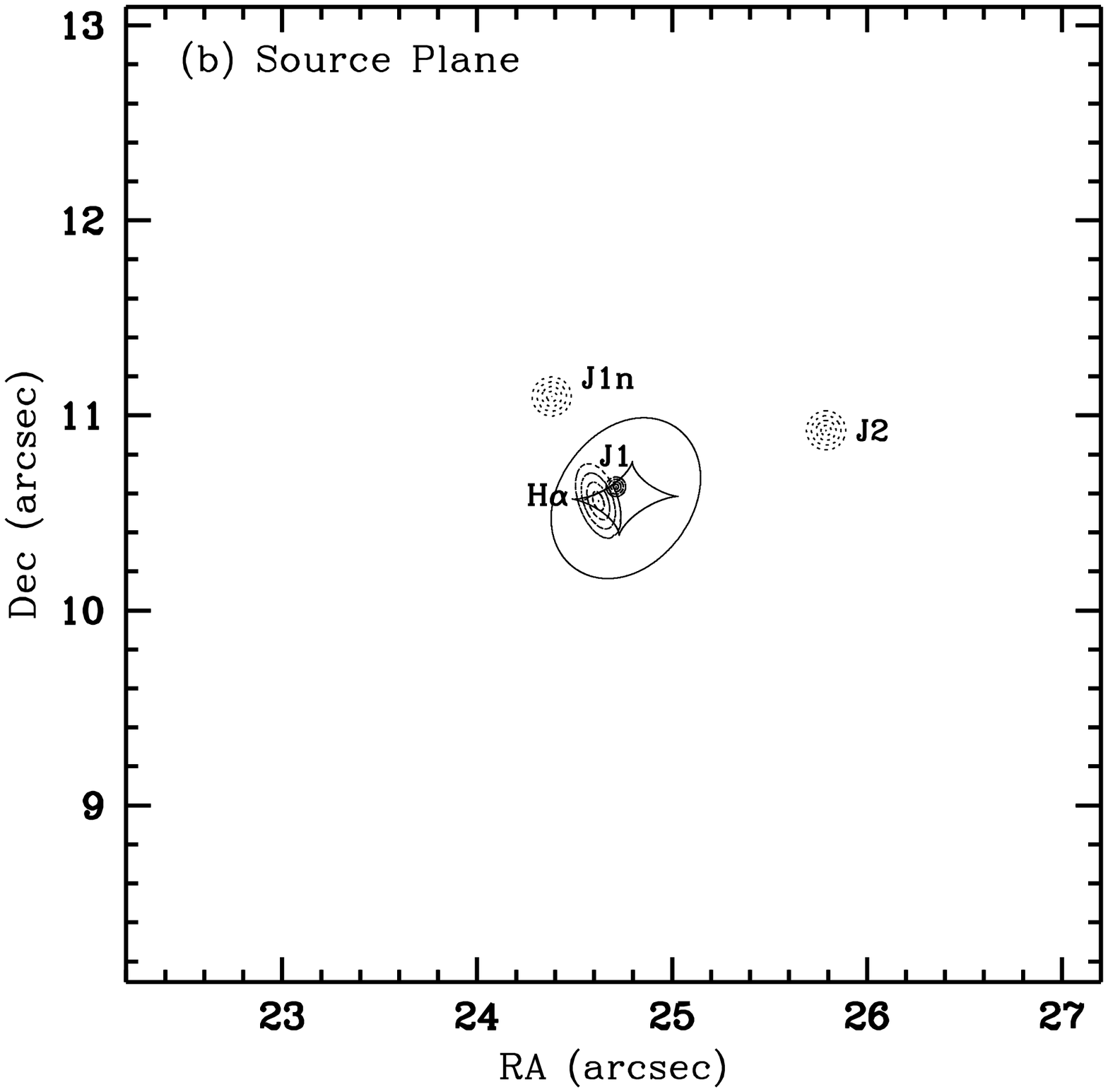}
\caption{The gravitational lensing model image of (a) the image plane overlaid on the \hst-$F702W$ image
and (b) the source plane.
The solid contours show the profile of J1 at z=2.565  with an intrinsic diameter of 0\farcs1 (0.8 kpc), 
while the dotted contours show that of J1n, J2, and the \ha~
emission line cloud.
Intrinsic sizes of J1n and J2 are 0\farcs2 (1.6 kpc), while that of
the \ha~cloud is an ellipse of 
0\farcs2$\times$0\farcs4 (1.6 kpc $\times$ 3.3 kpc).
}
\label{fig5}
\end{figure}

Figure \ref{fig5} shows the image- and source-plane of the model.
The solid contours show the profile of J1 at z=2.565 with an intrinsic diameter of 0\farcs1 (0.8 kpc).
It can be seen that J1a and J1b in the model image are elongated, 
while that in the \hst-$F702W$ image is not. 
This suggests that the intrinsic size of J1 is smaller than 0\farcs1 (0.8 kpc), 
and that what we are seeing is a kind of a circumnuclear starburst disk like that in a nearby ULIRG.
The total magnification factor of J1 is found to be 34$^{+9}_{-10}$.
However, we should keep in mind that this magnification factor is that of a point source. 
If the source is assumed to be extended by 0\farcs1, the total magnification factor 
becomes 26$^{+14}_{-3}$.

Model contours of J1n, J2, and the \ha~cloud are plotted with dotted lines.
The extended \ha~ emission line reported in \citet{tecza04} is well 
reproduced by a cloud with an intrinsic size of 0\farcs2$\times$0\farcs4 (1.6 kpc $\times$ 3.3 kpc)
located at 0\farcs1(0.8 kpc) west of J1. Its total magnification factor estimated from our model 
is 23$^{+4}_{-6}$ assuming that the \ha~emission line region is extended by 0\farcs3.
This value matches that of the CO emission which is estimated to be $\sim 25$ \citep{downes03}.

J1e, a counter image of J1 with half the flux of J1d, as well as \ha -d should appear in the 
west side of J1c.
However, both are not detected in either the \hst-$F702W$ or the \ha~image. 
There are several explanations for this that; (1) they are absorbed by dust in the line of sight, 
presumably in J1c, (2) the real potential of J1c deviates from the model and the flux of J1e and \ha -d
are depressed, and (3) J1a, J1b, and J1d are not a multiple image of
a single source, but different components of a merging source structure.
In the last case, the mass of J1c should be much smaller to avoid the multiple lensing.
However, because the optical properties of J1c well matches our model as discussed in section 4.2.2, 
it is unlikely that the mass of the J1c is much smaller than our model.
We therefore assume that J1e and \ha -d are depressed either by dust or 
by shape of the real potential.

\subsubsection{Properties of the Lensing Galaxy}
To check the feasibility of the gravitational lensing model, we 
next assess the optical properties of the lensing galaxy J1c and 
compare them with the model.

The profile of J1c in the \hst-$F702W$ image  is fitted 
with the \citet{sersic68} profile 
\begin{displaymath}
I(r)=I_e \exp \left[ -b_n \left \{ \left(\frac{r}{r_e}\right)^{1/n}-1  \right \} \right],
\end{displaymath}
where $b_n=1.9992n-0.3271$ \citep{graham01}.
Fitted parameters are $n=2.1$ and $r_e=0$\farcs5, and the 
flux within $r_e$ is estimated to be 3.1 $\mu$Jy.
Using spectra of 2--5 Gyr instantaneous burst model
calculated by PEGASE \citep{fioc97},
the stellar mass is then estimated to be 1.5--3.8$\times10^9\,M_{\odot}$.

On the other hand, the projected mass within $r_e$ of J1c estimated from
the gravitational lensing model is 
3.8$\times10^9\,M_{\odot}$ using the equation of \citet{kochanek95}, which contains 
both stellar mass and dark matter halo.
Considering the dark matter fraction in early-type galaxies of $\lesssim 0.5$ \citep{treu04},
this is a good agreement with the value obtained above and our gravitational lensing model seems 
reasonable.
In addition, the size ($r_e=$0\farcs5 $\sim$ 2 kpc) and the velocity dispersion
($\sigma=$123 km/s) indicate that J1c is a relatively small elliptical in Abell 1835.

\subsection{SEDs and Model Fitting}
To quantify the stellar contents of J1 and J2 together with their star formation history, 
we combine the optical $BRI$ photometry of I00 with the current spectra to 
construct SEDs from the optical to NIR.
Because I00 use 3\arcsec aperture for the photometry while our slit size is much smaller,
we measured the ratio of the flux between the different aperture sizes in the $R$-band, 
using the $HST$-$F702W$ image convolved to the 0\farcs7 seeing size 
of the $JH$-band spectroscopy. 
This ratio is used to correct all the fluxes of the optical bands in I00.
Also, an additional error of 0.1 mag is added to the optical data considering the 
systematic uncertainties in the flux calibration of the NIR spectra.
The emission lines in the NIR spectra are removed by the Gaussian fitting described in section 3.
The obtained SEDs are shown in Figure \ref{fig6}.
Continuum breaks at rest-frame 4000\AA\ are seen in both J1 and J2.

\subsubsection{Component J2}
We fit the SED of J2 by spectrum templates calculated by PEGASE \citep{fioc97}
to which an empirical dust extinction law of starburst galaxies by \citet{calzetti00} is applied.
Free parameters are dust extinction $E(B-V)$, age, and total mass.
The initial mass function of Salpeter is adopted.
We consider two extreme cases of star-formation history; one is an instantaneous burst, 
and the other is a continuous burst with duration of 1 Gyr.
The results are summarized in Table \ref{tab2}, and the fitted spectra are
shown in Figure \ref{fig6}b.
J2 is expected to be dominated by very young population,
and have formed of order of  $\sim10^8$\msol\ stars.
We must keep in mind that the SED is extracted from a small aperture
(1\arcsec$\times$2\arcsec ), 
and what we see is only the central region of J2.
3\arcsec aperture flux in the $K$-band (I00) indicates that
the obtained mass may be all underestimated by a factor of $\sim 3$.

\subsubsection{Component J1}
\begin{figure}
\figurenum{6}
\epsscale{0.7}
\plotone{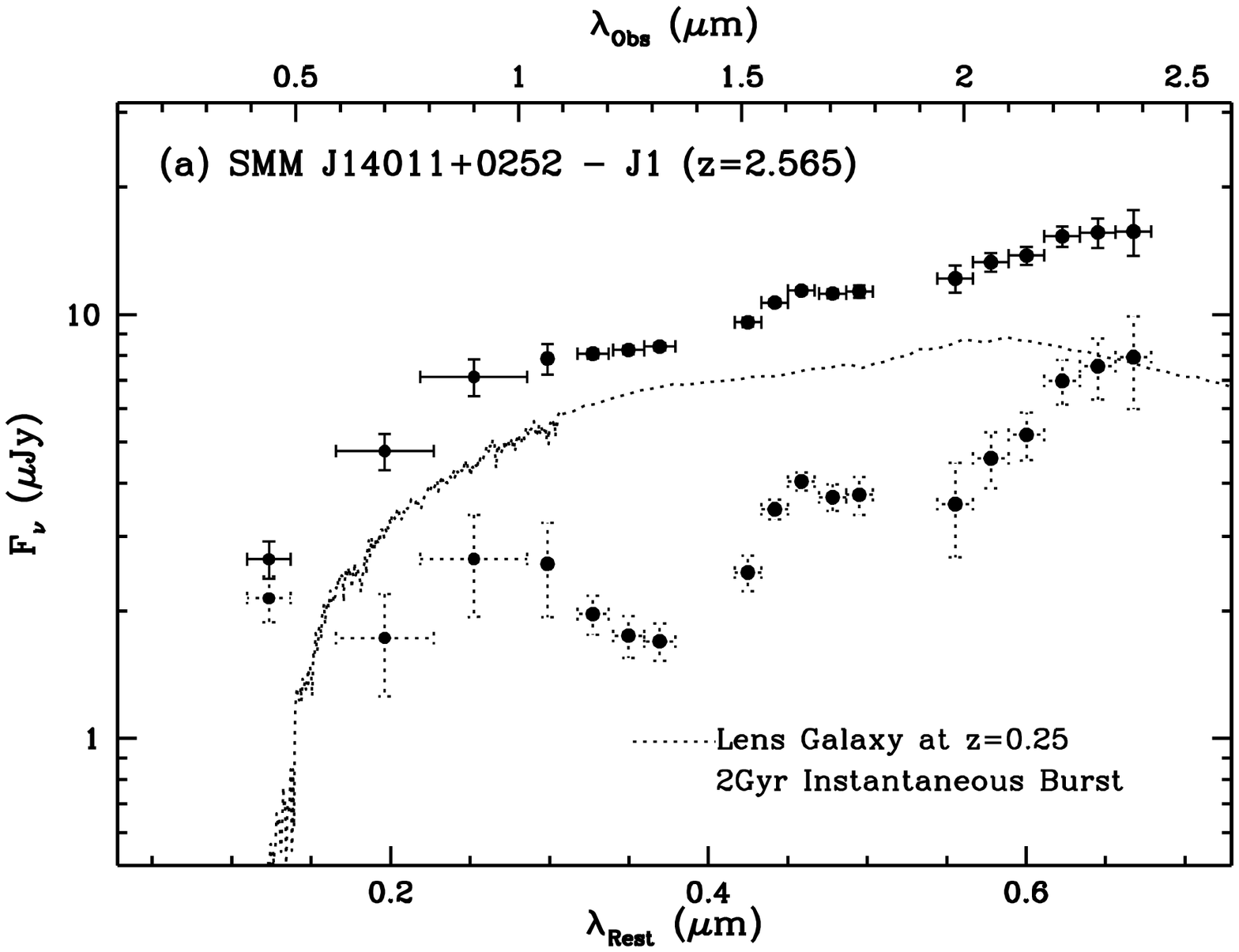}
\plotone{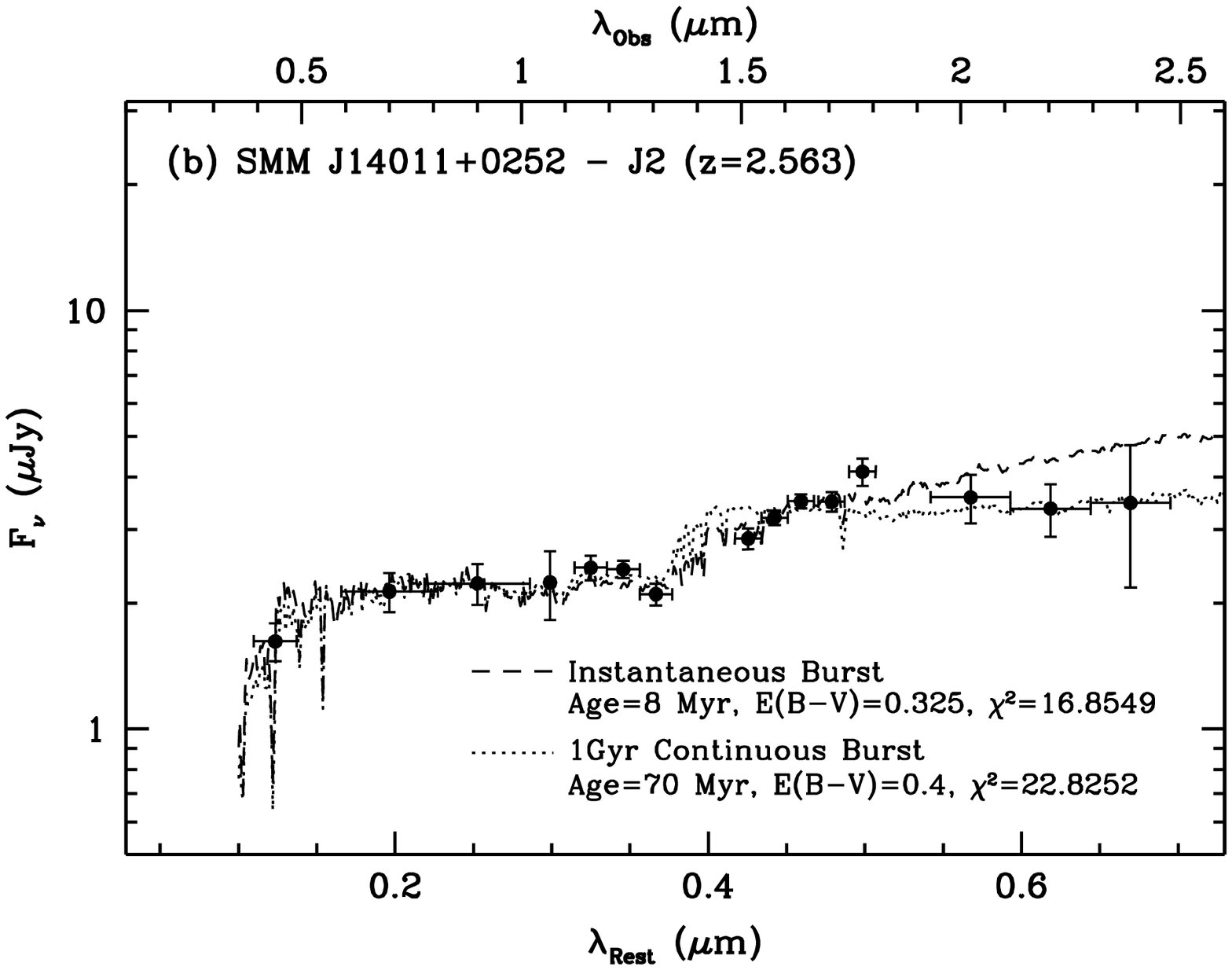}
\caption{Optical to NIR SEDs of (a) J1 and  (b) J2. Present data are shown as open circles, while the data taken from I00 are shown as
filled circles. (a) The dotted line shows the assumed model spectrum of
 $z=0.25$ lensing galaxy, and the dotted plots show the lens-subtracted
 SED.
(b) The dashed and dotted lines show the result of the fitting by instantaneous burst and 1 Gyr
 continuous burst models, respectively.}
\label{fig6}
\end{figure}

To assess the contamination of J1c to the SED of J1, 
the spectral shape of J1c is assumed to be that of a 2Gyr instantaneous burst model.
It is scaled by the flux obtained from the \hst-$F702W$ image,
and plotted in Figure \ref{fig6}a as the dotted line.
This model spectrum of J1c is then subtracted 
from the SED of J1 to obtain the intrinsic SED, which is shown as the dotted plots in Figure \ref{fig6}a.

It can be seen that the $B$-band flux is little contaminated by J1c.
This agrees with the result of I00 that 
the peak of J1 in the $U$-band is slightly offset toward east from that in the $R$-band, 
indicating that J1 at $z=2.565$ dominates the $U$- and $B$-band flux.

We next estimate the total intrinsic flux of J1 in the $K$-band to be 25 $\mu$Jy, 
by subtracting the model spectrum of J1c obtained above from the $K$-band flux of I00 (3\arcsec aperture).
Assuming the magnification factor of J1 to be 5 as a lower limit and 
the shape of the SED to be same as that of J2, the upper limit of the stellar
mass in J1 is estimated to be $\sim10^{9} M_{\odot}$. If more dust exists in this system than in J2, 
this upper limit of the stellar mass will become larger.

\subsubsection{Formation Site of a Normal Galaxy?}
The amount of gas reservoir in J1 is few$\times 10^{9}$\msol~, estimated from the CO observations 
\citep{downes03}.
Therefore, if SMM J14011+0252 is a place where a massive galaxy of $\sim 10^{12}\,M_{\odot}$ is being assembled, 
stellar mass of $\sim 10^{12}\,M_{\odot}$ must have already been assembled and hidden in dust. 
This is more than 100 times as much as the value calculated above,
and extinction of $A_V>5$ is required among the galaxy.
Although such an amount of extinction is broadly observed in the circumnuclear starburst disk 
in nearby ULIRGs \citep[e.g. ][]{genzel98},
their size is smaller ($<0.5$kpc) and
their stellar and gas masses are comparable to each other and much smaller 
($\sim 10^{10}\,M_{\odot}$) than those required in SMM J14011+0252.
Therefore, considering the difficulty to hide a stellar component of $10^{12}\,M_{\odot}$ in dust,
it is more probable that SMM J14011+0252 is a site of less massive 
($\lesssim 10^{10}\,M_{\odot}$) galaxy formation.

If we assume that the submillimeter radiation is emitted from the same 
region that emits CO or \ha~and its magnification factor is of order of 30, the intrinsic 
850 $\mu$m flux is $\sim 0.5$mJy.
This is the flux that most of less massive high-$z$ galaxies such as Lyman break galaxies 
are expected to emit at most \citep{chapman00} .
Also, the luminosity in the far-IR wavelength becomes $\sim10^{12}\,L_{\odot}$, which are same as 
that of local ULIRGs.

However, readers should keep in mind that above discussions 
are based on the uncertain gravitational lensing model, 
which does not have redshift information of the observed components to settle the membership.
The uncertainty in the model parameters are also large and the calculated total magnification
may change by factor of 2.
Also, there is a possibility that that the position of the 
submillimeter source is offset from that of CO or \ha~
and it does not suffer a strong lensing effect. 
In such case, the total magnification is only a factor of 5, and the intrinsic
luminosity in far-IR becomes an order of magnitude larger. Then, this
object may be interpreted as a massive and dusty starburst forming a
larger mass ($\sim 10^{11} M_{\odot}$) of stars.

It is likely that the starburst in this system are triggered by an encounter of J1 and J2.
If we ignore the radial distance (that is, assuming the distance between J1 and J2 is that observed in the source plane
of 1\farcs2=10 kpc)
and assume the encounter to have occurred 50 Myr ago,
the projected relative velocity is estimated to be 200 km s$^{-1}$. 
Considering the radial distance and motion of J1 and J2,
this matches the typical relative velocity of 1000 km s$^{-1}$ in a cluster of galaxies.

Therefore, J1 may be a high-$z$ counterpart of the local ULIRGs such as Arp 220, 
and its submillimeter flux are magnified above the detection limit of SCUBA by the strong lensing effect of J1c.
Its total barionic mass is $\sim 10^{10}\,M_{\odot}$, and 
we hypothesize that this object is a less-massive forming galaxy at $z\sim3$ which 
will evolve into a mid-sized elliptical or spiral galaxy such as the Milky-Way.

\section{Summary}
We have conducted $JHK$ spectroscopy of components J1 and J2 of 
the gravitationally lensed submillimeter galaxy SMM J14011+0252 at $z=2.565$ using 
OHS/CISCO at the Subaru telescope.

\ha, \hb, \oii\ lines in both J1 and J2 are detected , and \nii$\lambda\lambda6548,6583$ lines are also detected 
in J1. No \oiii\ lines are found. From the emission line diagnosis, the emission line cloud of J1 
is shown to be ionized/excited by stellar UV radiation, not by an AGN.
The continuum is detected in both J1 and J2, and 
the Balmer breaks are clearly seen. The SED of J2 is fitted by 
a model spectrum of young ($\sim50$ Myr) stellar population with
some amount of dust extinction.

The new gravitational lensing model is constructed to reproduce the small knots of J1,
which also reproduces the morphology of the extended \ha~ cloud well.
This model suggests that J1 may be strongly magnified by the elliptical galaxy in 
Abell 1835 by a factor of 30. 
Also, using this model together with the profile fitting of the lensing galaxy, 
the stellar mass of J1 at $z=2.565$ is estimated to be $\lesssim 10^{9}\,M_{\odot}$.
This indicates that SMM J14011+0252 is not a formation site of a present-day
massive elliptical, but more like a high-$z$ version of an ULIRG, 
and will become a present day spiral or mid-sized elliptical of
$10^{10-11} M_{\odot}$.

\acknowledgments
We appreciate M. Fioc and B. Rocca-Volmerange for generously offering their 
galaxy modeling code, PEGASE.
We thank C. Keeton for providing us his gravitational lensing software GRAVLENS, and
some instruction to use it.
We also thank I. Smail for providing us the information about the 
strong lensing of SMM J14011+0252, and the anonymous referee who gave us instructive advice to
refine this paper.
Part of the presented data were obtained during the test observation of OHS 
using the Subaru telescope, and we are indebted to all the staff of the Subaru 
telescope.
$HST$ data presented in this paper were obtained from the Multimission 
Archive at the Space Telescope Science Institute (MAST). STScI is operated by 
the Association of Universities for Research in Astronomy, Inc., under NASA 
contract NAS5-26555. Support for MAST for non-\hst\ data are provided by the 
NASA Office of Space Science via grant NAG5-7584 and by other grants and contracts.

\clearpage

\clearpage

\begin{table}
\begin{center}
\begin{tabular}{ccccc}
\tableline
\tableline
Target & Date & Band & $t_{exp}$ & PA\\
 & & & (s) & (\degree)\\
\tableline
J1/J2 & 2002 Feb 28 & $K$ & 4800 & $-69$ \\
J1/J2 & 2002 Mar 1 & $JH$ & 6000 & $-69$ \\
J1/J2 & 2002 Mar 2 & $JH$ & 4000 & $-69$ \\
J1/J1n & 2002 May 18 & $JH$ &8000 & $24$ \\
\tableline
\end{tabular}
\end{center}
\caption{Log of spectroscopic observations.\label{tab00}}
\end{table}

\begin{table}
\begin{center}
\begin{tabular}{cccc}
\tableline
\tableline
Component & Aperture & $F_{\lambda, K^{\prime}}$ & $F_{\lambda, H}$\\
&&($\times10^{-18}$W m$^{-2}$ \micron$^{-1}$) &($\times10^{-18}$W m$^{-2}$ \micron$^{-1}$) \\
\tableline
J1 & 0\farcs95$\times$2\farcs2 & 9.2$\pm$0.5 & 12.2$\pm$0.7 \\
J2 & 0\farcs95$\times$1\farcs7 & 2.4$\pm$0.4 & 3.8$\pm$0.5 \\
\tableline
\end{tabular}
\end{center}
\caption{Flux in the \kp\- and $H$-band of J1 and J2.\label{tab0}}
\end{table}

\begin{table}
\begin{center}
\begin{tabular}{ccccccc}
\tableline
\tableline
&Line &$\lambda_{\rm rest}$& Line Flux & Continuum Flux& EW$_{\rm rest}$ & $z$\\
& &(\AA)& ($\times10^{-19}$ W m$^{-2}$) & ($\times10^{-18}$ W m$^{-2}$ \micron$^{-1}$) & (\AA) \\
\tableline
J1 &\ha & 6563 & 1.3$\pm$0.4 & 8.9$\pm$1.8 & 42 & 2.565$\pm$0.001 \\
   &\nii& 6583 & 0.62$\pm$0.39 & 8.9$\pm$1.8 & 20 & 2.565$\pm$0.001 \\
   &\hb & 4861 & 0.33$\pm$0.12 & 10.9$\pm$0.5 & 8.5 & 2.562$\pm$0.002 \\
   &\oii& 3727 & 0.41$\pm$0.13 & 14.0$\pm$0.6 & 8.1 & 2.563$\pm$0.003 \\
\tableline
J2 &\ha & 6563 & 0.48$\pm$0.34 & 1.9$\pm$1.1 & 72 & 2.562$\pm$0.003 \\
   &\nii& 6583 & 0.27$\pm$0.33 & 1.9$\pm$1.1 & 40 & 2.562$\pm$0.003 \\
   &\hb & 4861 & 0.11$\pm$0.09 & 3.5$\pm$0.4 & 9.0 & 2.561$\pm$0.005 \\
   &\oii& 3727 & 0.26$\pm$0.09 & 3.6$\pm$0.4 & 20 & 2.564$\pm$0.003 \\
\tableline
\end{tabular}
\end{center}
\caption{Fitted parameters of the emission lines in J1 and J2.\label{tab1}}
\end{table}

\begin{table}
\begin{center}
\begin{tabular}{lccl}
\tableline
\tableline
Component &\multicolumn{2}{c}{Parameter}  & Comment\\
\tableline
J1c & $\sigma$ & 123$^{+10}_{-14}$ km/s & Velocity dispersion of the potential\\
& $e$ & 0.67$^{+0.09}_{-0.10}$ & Ellipticity of the potential\\
& $\theta_{e}$ &53$^{+9}_{-10}$ $^{\circ}$& Position angle of ellipticity\\
& $b$ & 0$^{+0.07}$  \arcsec & Core radius\\
\tableline
Abell 1835 & $r_s$ & 0.64$^{+5}_{-10}$ Mpc \\
& $\kappa_s$ & 0.189$^{+0.005}_{-0.015}$ \\
& $dx$ & 44$^{+7}_{-4}$ \arcsec & Distance from J1c in R.A. \\
& $dy$ & 20$^{+6}_{-4}$ \arcsec & Distance from J1c in Dec.\\
\tableline
\end{tabular}
\end{center}
\caption{Model parameters of the lensing potential.\label{tab11}}
\end{table}

\begin{table}
\begin{center}
\begin{tabular}{cc|ccc}
\tableline
\tableline
Source Name & Position\tablenotemark{a} & Image Name & Magnification & Position\tablenotemark{b}\\
\tableline
J1&$(24.71,~ 10.64)$ & J1a & 11  & $(-0.59,~ -0.18)$ \\
  & & J1b & 13  & $(-0.25,~ -0.67)$\\
  & & J1d & 6.3 & $(-0.23,~ 1.10)$\\
  & & (J1e) & 3.1 & $(0.46,~ -0.21)$\\
\tableline
J1n& $(24.38,~ 11.10)$ & J1n & 4.4 & $(-1.00,~ 2.25)$ \\
\tableline
J2& $(25.79,~ 10.93)$& J2 & 5.7 & $(1.89,~ 0.80)$ \\
\tableline
\ha& $(24.62,~ 10.57)$ & \ha -a & 12 & $(-0.78,~ 0.28)$\\
  & & \ha -b& 10 &$(-0.44,~ -0.73)$\\
  & & \ha -c& 9.5&$(-0.51,~ 0.90)$ \\
  & & (\ha -d) & 2.0 & $(0.36,~ -0.06)$\\
\tableline
\end{tabular}
\end{center}
\tablenotetext{a}{Position on the source plane in arcsec, relative to J1c.}
\tablenotetext{b}{Position on the image plane in arcsec, relative to J1c.}

\caption{Calculated parameters of the lensed images. The position of each components is shown in 
Figure \ref{fig5}. The parenthesized images are those not detected.
Note that these magnification factors are calculated asuuming that a lensed image is a point source.
If the image is extended, the magnification factor becomes smaller (see text).
\label{tab12}}
\end{table}

\begin{table}
\begin{center}
\begin{tabular}{ccccccc}
\tableline
\tableline
Model & Age & $E(B-V)$ & Stellar Mass & Total Mass & SFR& $\chi^2$  \\
 & (Myr) & & (\msol) & (\msol) & (\msol\ yr$^{-1}$) & \\
\tableline
 Instantaneous & 8 &0.33& 6.1$\times10^7$ & 6.2$\times10^{7}$ & 0.0  &16.85\\
 1Gyr Continuous & 70 &0.40& 1.7$\times10^8$ & 2.4$\times10^{9}$ & 4.4  &22.83\\
\tableline
\end{tabular}
\end{center}
\caption{Parameters for the best-fit models of the J2 SED. Mass and SFRs are corrected for 
the gravitational lensing effect with the magnification factor of 5.7.\label{tab2}}
\end{table}

\end{document}